\documentclass{desyproc}

\newcommand{\etal}{{\it et al.}}

\newcommand{\eqsal}[1]{\begin{align}#1\end{align}}
\newcommand{\ce}[1]{Eq.~(\ref{#1})}

\newcommand{\cf}[1]{{Fig.~\ref{#1}}}

\newcommand{\I}{{\cal I}}

\newcommand{\beq}[1]{
\begin{equation}\label{#1}}
\newcommand{\eeq}{\end{equation}}

\newcommand{\re}{\mathrm{Re}\,}
\newcommand{\im}{\mathrm{Im}\,}

\begin{document}

\title{
\begin{flushright}
\rm \footnotesize
SLAC-PUB-13726\\
CPHT-PC075.0709 
\end{flushright}
On the physical Relevance of the Study of \\ $\gamma^\star \gamma \to \pi^0 \pi ^0$ at small $t$ and large $Q^2$}

\author{{\slshape J.P. Lansberg$^1$, B. Pire$^2$, L. Szymanowski$^3$}\\[1ex]
$^1$SLAC National Accelerator Laboratory, Theoretical Physics, Stanford University,\\ Menlo Park, CA 94025, USA\\[1ex]
$^2$Centre de Physique Th\'eorique, \'Ecole Polytechnique - CNRS, 91128 Palaiseau, France\\[1ex]
$^3$Soltan Institute for Nuclear Studies, Warsaw, Poland
}



\maketitle

\begin{abstract}
We discuss the relevance of a dedicated measurement of exclusive production of a pair
of neutral pions in a hard $\gamma^\star \gamma$ scattering at small momentum transfer.
In this case, the virtuality of one photon provides us with a hard scale in the process,
enabling us to perform a QCD calculation of this reaction rate using 
the concept of Transition Distribution Amplitudes (TDA). Those are
related by sum rules to the pion axial form factor $F^{\pi}_A$ and,
as a direct consequence, a cross-section measurement of this process 
at intense beam electron-positron colliders such as CLEO, KEK-B and PEP-II, or Super-B
would provide us with a unique measurement of the neutral pion axial form factor $F^{\pi^0}_A$ at small scale.
\end{abstract}

\section{Introduction}

In a series of  papers \cite{Pire:2004ie,Pire:2005ax,Pire:2005dh,Lansberg:2006fv,Lansberg:2006uh,Lansberg:2007ec,Lansberg:2007se,TDAreview}, we have advocated that  factorisation theorems \cite{fact} 
for exclusive processes may be extended to the case of other reactions such as ($M_i$ stands for
a meson and $B_i$ for a baryon)
$B_1\,\overline{B}_2 \, \to \,\gamma^\star\,\gamma$, $B_1\,\overline{B}_2 \, \to \,\gamma^\star\,M_1$ 
$\gamma_T^\star\,B_1 \to \,  B_2 \gamma$,  $\gamma_T^\star\,B_1 \to \,  B_2 M_1$ or $\gamma_L^\star\,\gamma \to \, M_1 M_2$,
in the kinematical regime where the off-shell photon is highly virtual ($Q^2$ of 
the order of the energy squared of the reaction) but the momentum transfer $t$
is small. This enlarges the successful description of deep-exclusive $\gamma \gamma$  
reactions  in  terms   of  distribution  amplitudes \cite{ERBL}  and/or
generalised  distribution amplitudes \cite{GDAAPT}  on the  one  side and
perturbatively   calculable  coefficient  functions   describing  hard
scattering at  the partonic  level on the  other side.  

Intense beam electron colliders, such as $B$ factories, are ideal places to study such reactions as
   \begin{equation} 
  \gamma_L^\star\,\gamma \to  \rho^\pm\pi^\mp,~\gamma_L^\star\,~\gamma \to  \pi^\pm\pi^\mp,~ 
\gamma_L^\star\gamma \to \pi^0\pi^0, \nonumber
  \label{gagapi}
\end{equation} 
in the  near forward region and for large virtual photon invariant mass 
$Q$. Recently {\sc BaBar} reported a new measurement of  the reaction $\gamma^\star \gamma \to \pi^0$ 
up to photon virtualities squared of 40~GeV$^2$~\cite{Aubert:2009mc}.
In the latter study, the reaction $\gamma^\star \gamma \to \pi^0 \pi ^0$ was investigated 
in the $f_2(1270)$ and $f_0(980)$ resonance region
as a potential background for the study of the $\pi^0$ transition form factor.
This low-$W_{\pi\pi}^2$ kinematical region should be analysed in the framework of generalised two-meson distribution
 amplitudes~\cite{GDAAPT} and in particular should solve the much discussed problem of its phase structure around 
the $f_0$ mass \cite{DI} which is of crucial importance for the ability to detect Pomeron-Odderon 
interference effects in high energy electro-production of meson pairs \cite{HPST}.

We want here to emphasise another kinematical region, namely the small-$t$ large-$W_{\pi\pi}^2$ region at moderate 
$Q^2$ (2 GeV$^2$ and more) which may provide us with unique information
on the $\pi^0$ axial form factor at small scale which so far has never been experimentally measured. 
It has been argued that  a new duality~\cite{Anikin:2008bq} relates these two factorisation regimes.

In principle, another possibility to study this quantity would be the crossed channel, that is DVCS 
on a virtual neutral pion along the lines exposed in Ref.~\cite{Amrath:2008vx} for $\pi^+$.

\section{Pion-pair production in the TDA regime}

 Let us recall the main ingredients of the analyses developed in \cite{Pire:2004ie,Lansberg:2006fv} focusing on the 
neutral pion case. With the kinematics  described in \cf{fig:ggstarpipi}, we define the axial 
$\gamma \to \pi$ transition distribution 
amplitude (TDA)  $A(x, \xi, t)$ as the Fourier transform of matrix element 
$\langle     \pi^0(p_\pi)|\, {\cal O}_A \,|\gamma(p_\gamma) \rangle$ where 
${\cal O}_A = \bar{\psi}(\frac{-z}{2})[\frac{-z}{2},\frac{z}{2}]\, \gamma^\mu \gamma^5{\psi}(\frac{z}{2})$.  
The Wilson line $[\frac{-z}{2},\frac{z}{2}]$  ensures  the QCD-gauge
invariance for  non-local operators  and equals unity in  a light-like
(axial) gauge. We do not write the  electromagnetic Wilson line, since 
we choose an electromagnetic axial gauge for the photon.
We then factorise the amplitude of the process  $  \gamma_L^\star \gamma \to \pi^0 \pi^0$
as
\begin{equation}
\label{}  
\sum_{q=u,d}\int dx dz \,\Phi^q_{\pi}(z)  M^q_{h}(z,x,\xi) \frac{A^{\pi^0}_q(x, \xi, t)}{f_\pi}\;,
\end{equation}
 with a hard amplitude $M^q_{h}(z,x, \xi)$, 
$\Phi^q_{\pi}(z)$  the  distribution amplitude (DA) 
for $q$ quark content of the $\pi$ meson with momentum $p'_\pi$ and  $A^{\pi^0}_q(x, \xi, t)$ the axial $\gamma \to\pi$ TDA for the quark $q$.

\begin{figure}[h]
\centering{\includegraphics[width=8cm]{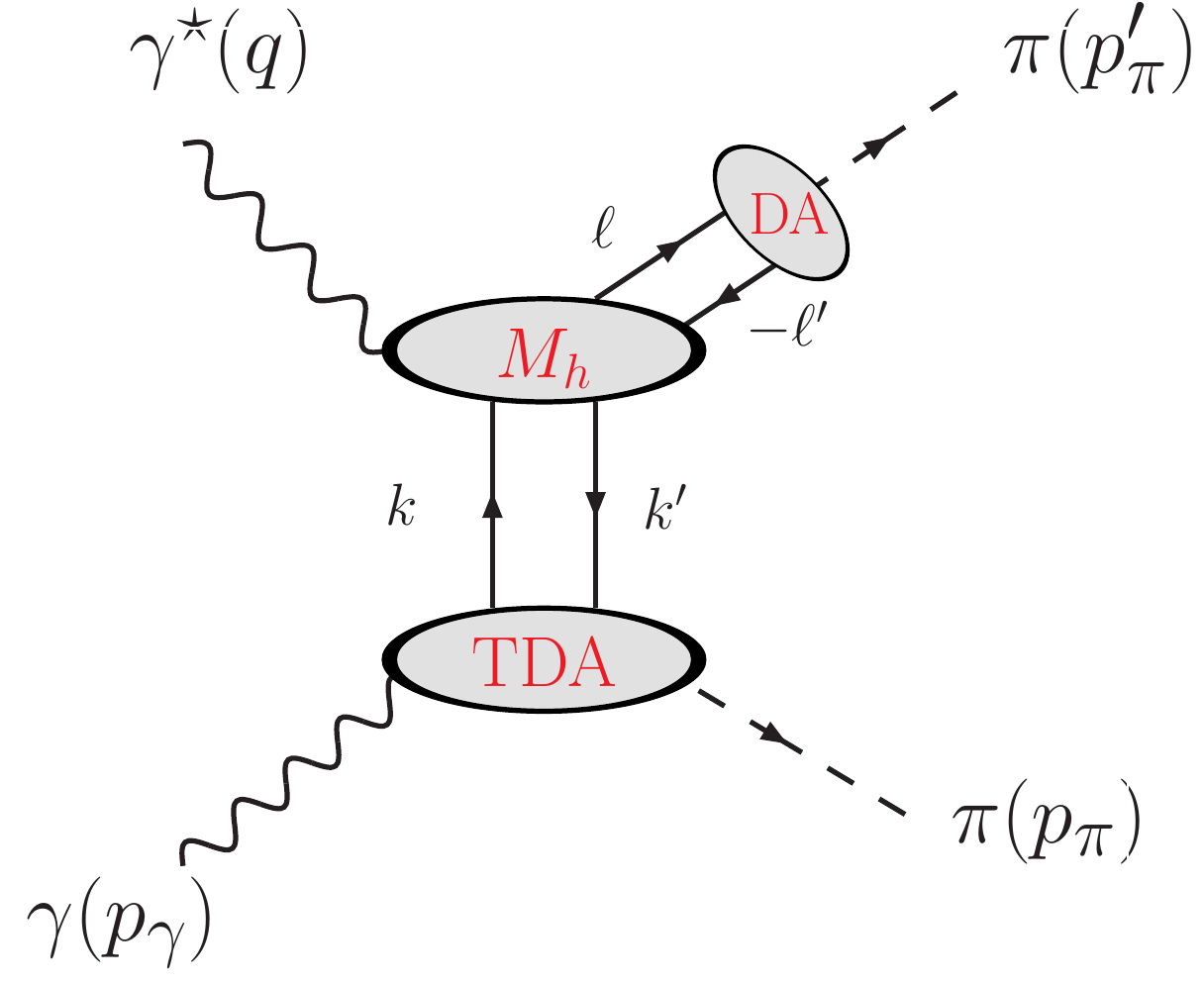}}
\caption{The factorised amplitude for $\gamma^\star \gamma \to \pi^0 \pi^0$ at small transfer
momentum.}
\label{fig:ggstarpipi}
\end{figure}

The variable $z$ is as usual the light-cone momentum fraction carried by the quark entering the
pion with momentum $p'_\pi$, $x+\xi$ (resp. $x-\xi$) is the corresponding one for the quark leaving (resp. entering) 
the TDA. The skewness variable $\xi$ describes the loss of light-cone momentum of the incident 
photon and is connected to the Bjorken variable $x_B$.

Contrarily to the case of generalised parton distributions (GPD) 
where the forward limit is related to the conventional parton distributions 
measured in the deep inelastic scattering (DIS), there is no such 
interesting constraints for the TDAs.
The constraints we have here are sum rules obtained by taking the local limit of
the corresponding operators and soft limits when the momentum
of the meson in the TDA vanishes.

Let us consider in more detail the $ \gamma  \to\pi^{0} $ axial
TDAs which is defined by ($P=\frac{p_{\pi}+p_\gamma}{2}$, 
$\Delta=p_{\pi}-p_\gamma$):

\begin{equation} 
 \int \frac{d z^-}{2\pi}e^{ix P^+ z^-} 
\langle  \pi^0| \bar{q}(\frac{-z}{2})\Big[\frac{-z}{2};\frac{z}{2}\Big]\gamma^\mu\gamma^5 q(\frac{z}{2})|
\gamma \rangle   =
\frac{1}{P^+} \frac{e}{f_\pi}(\varepsilon\cdot \Delta) P^\mu A_q^{\pi^0}(x,\xi,t)
\label{eq:TDA_A}
\end{equation}

 A sum rule  may  be derived  for  this  photon to meson
TDA by integrating on $x$ both side of \ce{eq:TDA_A} and we get 
 \begin{equation} 
\frac{\;e}{f_\pi}(\varepsilon\cdot \Delta) P^\mu  \int^1_{-1} dx \, \,A^{\pi^0}_q(x,\xi, t) =\langle  \pi^0| \bar{q}(0)\gamma^\mu\gamma^5 q(0)|
\gamma \rangle,
\label{srVpi0}
\end{equation}
The latter matrix element of a local quark--anti-quark operator is directly related to the quark $q$ contribution $F_{A,q}^{\pi^0}$ to the axial form factor of the $\pi^0$ meson.
Similarly, we have in the vector charged pion case~\cite{Lansberg:2006fv}:
\begin{equation} 
\int^1_{-1} dx \, V^{\pi^{\pm}}(x,\xi,t)= \frac{f_\pi}{m_\pi} \,F^{\pi^{\pm}}_V(t), 
\label{srAV}
\end{equation}
with $F_V^{\pi^{\pm}}=0.017  \pm 0.008$~\cite{PDG}.

This sum rule constrains possible parametrisations of the TDAs.
Note, in particular, the $\xi$-independence of the right hand side of the relation.

\section{Models and cross section evaluation}

\subsection{Amplitude}

Let us thus consider the $\pi^0 \pi^0$ production case  when the $\pi^0$ with
momentum $p'_\pi$ flies in the direction of the virtual photon and the other 
$\pi^0$ emerges from the TDA. For definiteness, we choose, in the CMS of the meson pair,  
$p=\frac{Q^2+W_{\pi\pi}^2}{2(1+\xi)W_{\pi\pi}}(1,0,0,-1)$ 
and $n=\frac{(1+\xi)W_{\pi\pi}}{2(Q^2+W_{\pi\pi}^2)}(1,0,0,1)$  and we express the momenta 
 trough a Sudakov decomposition (with $ \Delta_T^2=\frac{1-\xi}{1+\xi} t$ and neglecting the pion mass):
\begin{equation}
p_\gamma= (1+\xi) p,\ \ 
q= \frac{Q^2+W_{\pi\pi}^2}{1+\xi} n - \frac{Q^2}{Q^2+W_{\pi\pi}^2}(1+\xi) p,\ \
p_{\pi}=(1-\xi) p - \frac{\Delta_T^2}{1-\xi} n + \Delta_T.
\end{equation}

We can see that $\xi$ is determined by the external kinematics 
through $\xi\simeq\frac{Q^2}{Q^2 + 2W_{\pi\pi}^2}$ -- similarly to $x_B=\frac{Q^2}{Q^2 + W_{\pi\pi}^2}$ to which it
is linked via the simple relation $\xi\simeq\frac{x_B}{2-x_B}$.

The  hard amplitude amplitude in \ce{gagapi} thus reads :
\begin{eqnarray}
M^q_{h}(z,x,\xi) =  
\frac{4\,\pi^2\,\alpha_{em}\,\alpha_s\,C_F\, Q_q}{N_C\,Q}
\frac{1}{z\,\bar z}\left(\frac{1}{x-\xi+i\epsilon} + 
\frac{1}{x+\xi-i\epsilon}  \right)\varepsilon \cdot  \Delta\,,
\end{eqnarray}   
where $Q_u=2/3$, $Q_d=-1/3$ and with $\bar  z = 1-z$. Note that the factor $f_\pi$
in the $\pi$ DA $\Phi^q_\pi(z)$  cancels with the one from the TDA definition
and does not appear in \ce{eq:iAx}.  Now, if we choose the asymptotic form for the neutral
pion DA, $\Phi_{\pi^0}^u(z)=-\Phi_{\pi^0}^d(z)=\frac{6 f_\pi }{\sqrt{2}} z (1-z)$,
the $z$-integration is readily carried out and after separating 
the real and imaginary parts of the amplitude, the $x$-integration gives:
\eqsal{
\label{eq:iAx}
\I_x^A=\frac{1}{\sqrt{2}}\sum_{q=u,d} |Q_q|\int_{-1}^1& dx\left(\frac{1}{x-\xi+i\epsilon} + 
  \frac{1}{x+\xi-i\epsilon}  \right) A^{\pi^0}_q(x, \xi, t)=\\
\frac{1}{\sqrt{2}}\sum_{q=u,d} |Q_q|&\Bigg[ \int^{1}_{-1} dx\,\frac{A^{\pi^0}_q(x,\xi,t)-
A^{\pi^0}_q(\xi,\xi,t)}{x-\xi}+ A^{\pi^0}_q(\xi,\xi,t) (\log\left(\frac{1-\xi}{1+\xi}\right)-i\pi)+\nonumber\\
& \int^{1}_{-1} dx\, \frac{A^{\pi^0}_q(x,\xi,t)-A^{\pi^0}_q(-\xi,\xi,t)}{x+\xi}
+A^{\pi^0}_q(-\xi,\xi,t)(\log\left(\frac{1+\xi}{1-\xi}\right)+i\pi )\Bigg].\nonumber
}

The  scaling law for  the amplitude  is 
\begin{equation} 
{\cal M}^{TDA}_{\gamma^\star \gamma}(Q^2, \xi,t) \sim \frac{\alpha_s\sqrt{-t}}{Q}\;,
\label{scaling}
\end{equation} 
up to logarithmic corrections due to the anomalous dimension of the TDA and the running of $\alpha_s$.

\begin{figure}[h]
\centering\includegraphics[width=14cm]{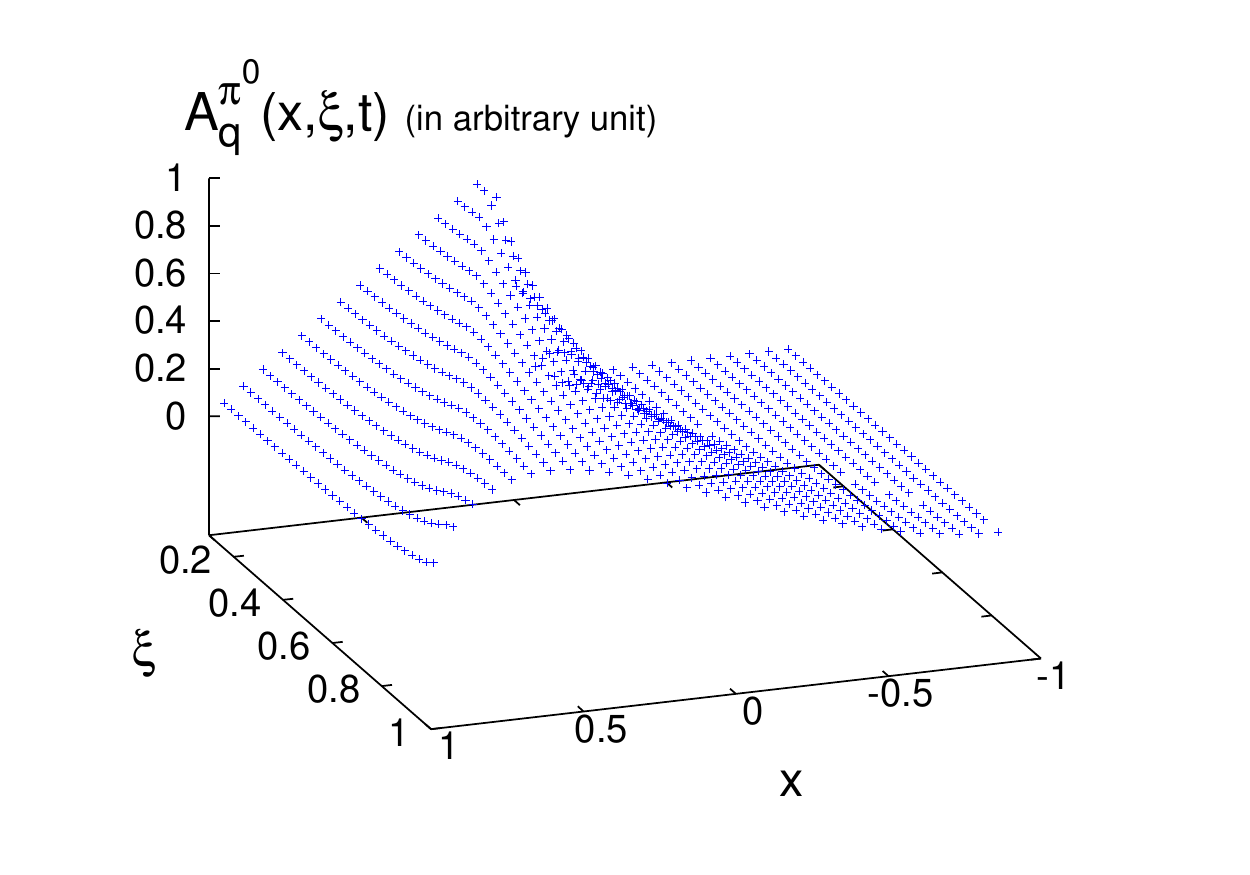}
\caption{The  $\gamma \to \pi^{0}$  axial transition distribution amplitude $A^{\pi{0}}_q(x,\xi,t)$ in Model 1 of Ref.~\cite{Lansberg:2006fv} (for $t=-0.5$ GeV$^2$) in arbitrary unit.}
\label{GPITDA}
\end{figure}

\subsection{Remarks on available models}

Lacking any non-perturbative calculations of
matrix element defining TDAs, we have initially built a toy model~\cite{Lansberg:2006fv} based on double distributions~\cite{rad} 
to get  estimates for the cross sections, to be compared with experimental data. In~\cite{Lansberg:2006fv}, we compared 
the rate obtained with this model with the one from the model built in~\cite{Tiburzi:2005nj}. Subsequently, a model
based on quark spectral representation  was developed in~\cite{Bro}, another based on NJL model
was studied in \cite{Courtoy:2007vy,Courtoy:2008nf} and lastly the $\pi \to \gamma$ TDAs were studied in a non-local chiral
  quark model~\cite{Kotko:2008gy}. All the models (see e.g.~\cite{GPD_pion}) used so far 
for the pion GPDs could be extended to the construction of $\pi \to \gamma$ TDAs. We refer to the different references 
for details. For illustration, we show here on \cf{GPITDA} 
the TDA $A(x, \xi, t)$ obtained in Ref.~\cite{Lansberg:2006fv} in arbitrary unit; its normalisation would be eventually fixed by the experimental data.

For the purpose of this note, we only need a rough  evaluation of the order of magnitude of the cross section
and will only use the Model 1 of Ref.~\cite{Lansberg:2006fv}. When a dedicated experimental analysis is being carried 
out, a careful survey of the cross sections obtained from the different 
theoretical models will be in order. Hence, based on a first experimental study of the $\xi$ dependence and after having
checked the scaling in $Q^2$,
we shall be in position to see which model describes best the physics involved. For this best model, 
we could then obtain by sum rules relations a first measurement of the axial $\pi^0$ form factor.

For the following, we shall show results for 
$\langle \pi^0| \bar d {\cal O}_A d |  \gamma  \rangle=-1/2 \langle\pi^0| \bar u {\cal O}_A u |  \gamma \rangle$
expected from the different charges of the $u$ and $d$ quarks
and using (from isospin arguments)
\begin{equation}
\langle \pi^+ | \bar d {\cal O}_A u | \gamma \rangle =
\langle \pi^0 | \bar d {\cal O}_A d |\gamma \rangle - \langle \pi^0   | \bar u {\cal O}_A u |\gamma \rangle.
\end{equation}
This would give $A^{\pi^{0}}_d=1/3 A^{\pi^+}$ and  $A^{\pi^{0}}_u=-2/3 A^{\pi^+}$. Note that more realistic models 
may give significantly larger rates.

\begin{figure}[!hbt]
\centering\includegraphics[width=10cm]{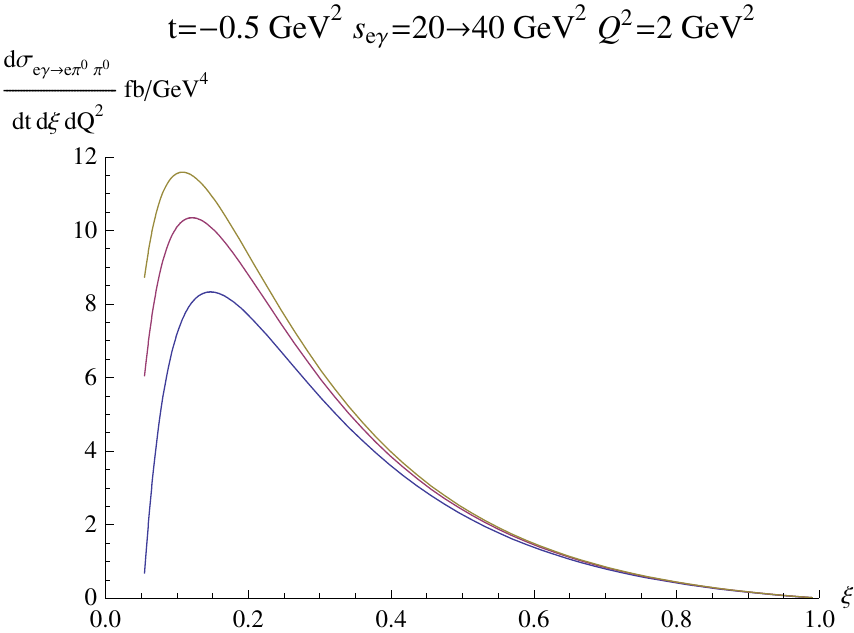}
\caption{Differential cross sections $\frac{d\sigma_{e\gamma \to e \pi^0 \pi^0}}{dQ^2 dt  d\xi}$ for the TDA  subprocess
as a function of $\xi$ for $Q^2=2$ GeV$^2$, $t=-0.5$ GeV$^2$ and 3 values of $s_{e\gamma}$: 20, 30 and 40 GeV$^2$ (from bottom to top). 
}
\label{fig:ds_pi}
\end{figure}

\subsection{Cross section}

Taking into account the contribution from the fermionic line for the emission by the electron of a longitudinal photon,
averaging over the real photon polarisation and integrating over $\varphi$ thanks to the 
$\varphi$-independence
of the TDA process, we eventually obtain the differential cross section\footnote{A factor $1/4$ is missing in Eq.(23) 
  of~\cite{Lansberg:2006fv}.}:
\beq{dsdxidtdq2}
\frac{d\sigma_{e\gamma \to e \pi^0 \pi^0}^{TDA}}{dQ^2 dt  d\xi}=
\frac{64 \pi \alpha_{em}^3 \alpha_s^2  2\pi }{9 (\xi + 1)^4 Q^8 }
(- 2  \xi t)(1-  \xi- (1+ \xi) \frac{W_{\pi\pi}^2}{s_{e\gamma}})  (\re^2(\I^A_x)+\im^2(\I^A_x)).
\eeq
 For the hypothesis discussed above, the resulting
cross section is roughly one sixth of the one obtained in~\cite{Lansberg:2006fv} for the charged case. The evolution as function of $\xi$
is displayed on \cf{fig:ds_pi}. Note that for small $\xi$ (particularly $W_{\pi\pi}^2\to Q^2$), the cross section shows a peak.

The $Q^2$-behaviour is model independent
and thus constitutes a crucial test of the validity of our approach.

\section*{Conclusion}

 We believe that our models for the photon to meson transition
distribution amplitudes are sufficiently constrained to give reasonable orders of magnitude for
the estimated cross sections. Cross sections are large enough for quantitative
studies to be performed at high luminosity $e^+e^-$ colliders. After verifying the scaling and the $\varphi$ independence
of the cross section, one should be able to measure these new hadronic matrix elements, and thus 
open a new gate to the understanding of the hadronic structure. 
In particular, we argued here that the study of $\gamma^\star \gamma \to \pi^0 \pi^0$ in the TDA regime could 
provide with a unique experimental measurement of the $\pi^0$ axial form factor. 
\section*{Acknowledgments}

We are thankful to I. Anikin, S.J. Brodsky, M. Davier, S. Li, S. Wallon for useful discussions and correspondence,
and the organisers of the Photon 2009 conference for their invitation.
This work is supported in part by a Francqui fellowship of the Belgian American Educational Foundation,
by the U.S. Department of Energy under contract number DE-AC02-76SF00515,  the French-Polish scientific agreement Polonium and
the Polish Grant N202 249235.


\begin{footnotesize}

\end{footnotesize}



\begin{thebibliography}{99}



\bibitem{Pire:2004ie}
  B.~Pire and L.~Szymanowski,
  Phys.\ Rev.\  D {\bf 71} (2005) 111501
  [arXiv:hep-ph/0411387].

\bibitem{Pire:2005ax}
  B.~Pire and L.~Szymanowski,
  Phys.\ Lett.\  B {\bf 622} (2005) 83
  [arXiv:hep-ph/0504255].


\bibitem{Pire:2005dh}
  B.~Pire and L.~Szymanowski,
  Acta Phys.\ Polon.\  B {\bf 37} (2006) 893
  [arXiv:hep-ph/0510161].


\bibitem{Lansberg:2006fv}
  J.~P.~Lansberg, B.~Pire and L.~Szymanowski,
  Phys.\ Rev.\  D {\bf 73} (2006) 074014
  [arXiv:hep-ph/0602195].

\bibitem{Lansberg:2006uh}
  J.~P.~Lansberg, B.~Pire and L.~Szymanowski,
  Nucl.\ Phys.\  A {\bf 782} (2007) 16
  [arXiv:hep-ph/0607130].

\bibitem{Lansberg:2007ec}
  J.~P.~Lansberg, B.~Pire and L.~Szymanowski,
  Phys.\ Rev.\  D {\bf 75}, 074004 (2007)
  [Erratum-ibid.\  D {\bf 77}, 019902 (2008)]
  [arXiv:hep-ph/0701125].


\bibitem{Lansberg:2007se}
  J.~P.~Lansberg, B.~Pire and L.~Szymanowski,
  Phys.\ Rev.\  D {\bf 76}, (2007) 111502R 
  [arXiv:0710.1267 [hep-ph]].


\bibitem{TDAreview}
For recent mini-reviews: J.~P.~Lansberg, B.~Pire and L.~Szymanowski,
In ''Exclusive Reactions at High Momentum Transfer'' (Singapore, World Scientific, 2008,  p. 367)
[0709.2567 [hep-ph]];
  J.~P.~Lansberg, B.~Pire and L.~Szymanowski,
  Nucl.\ Phys.\ Proc.\ Suppl.\  {\bf 184}, 239 (2008)
  [arXiv:0710.1294 [hep-ph]].


\bibitem{fact} 
  J.~C.~Collins, L.~Frankfurt and M.~Strikman,
  Phys.\ Rev.\ D {\bf 56} (1997) 2982.


\bibitem{ERBL} 
  A.~V.~Efremov and A.~V.~Radyushkin,
  Phys.\ Lett.\ B {\bf 94} (1980) 245;
  G.~P.~Lepage and S.~J.~Brodsky,
  Phys.\ Lett.\ B {\bf 87} (1979) 359.


\bibitem{GDAAPT}
 D.~Mueller et al.,
  Fortsch.\ Phys.\  {\bf 42}, 101 (1994);
M.~Diehl et al.,
  Phys.\ Rev.\ Lett.\  {\bf 81}, 1782 (1998) and
  Phys.\ Rev.\ D {\bf 62}, 073014 (2000);
 B.~Pire and L.~Szymanowski,
  Phys.\ Lett.\ B {\bf 556}, 129 (2003);
I.~V.~Anikin et al.,
  Phys.\ Rev.\ D {\bf 69}, 014018 (2004) and Phys.\ Lett.\  B {\bf 626} (2005) 86.


\bibitem{Aubert:2009mc}
  B.~Aubert  [The BABAR Collaboration],
  arXiv:0905.4778 [hep-ex].


\bibitem{DI}
  N.~Warkentin, M.~Diehl, D.~Y.~Ivanov and A.~Schafer,
  Eur.\ Phys.\ J.\  A {\bf 32}, 273 (2007)
  [arXiv:hep-ph/0703148].


\bibitem{HPST}
  P.~Hagler et al.,
   Phys.\ Lett.\  B {\bf 535}, 117 (2002) and Eur.\ Phys.\ J.\  C {\bf 26}, 261 (2002)





\bibitem{Anikin:2008bq}
  I.~V.~Anikin, I.~O.~Cherednikov, N.~G.~Stefanis and O.~V.~Teryaev,
  Eur.\ Phys.\ J.\  C {\bf 61}, 357 (2009) and arXiv:0907.2579 [hep-ph].





\bibitem{Amrath:2008vx}
  D.~Amrath, M.~Diehl and J.~P.~Lansberg,
  Eur.\ Phys.\ J.\  C {\bf 58} (2008) 179
  [arXiv:0807.4474 [hep-ph]].



\bibitem{PDG}
  C. Amsler {\it et al.}  [Particle Data Group],
  Phys.\ Lett.\ B {\bf 667} (2008) 1.





 \bibitem{rad}
  A.~V.~Radyushkin,
  Phys.\ Rev.\ D {\bf 59} (1999) 014030.

 

\bibitem{Tiburzi:2005nj}
  B.~C.~Tiburzi,
  Phys.\ Rev.\ D {\bf 72} (2005) 094001.


\bibitem{Bro}
 W.~Broniowski and E.~R.~Arriola,
  Phys.\ Lett.\  B {\bf 649} (2007) 49;

\bibitem{Courtoy:2007vy}
  A.~Courtoy and S.~Noguera,
  Phys.\ Rev.\  D {\bf 76} (2007) 094026
  [arXiv:0707.3366 [hep-ph]].


\bibitem{Courtoy:2008nf}
  A.~Courtoy and S.~Noguera,
  Phys.\ Lett.\  B {\bf 675} (2009) 38
  [arXiv:0811.0550 [hep-ph]].

\bibitem{Kotko:2008gy}
  P.~Kotko and M.~Praszalowicz,
  arXiv:0803.2847 [hep-ph].





\bibitem{GPD_pion}
A.~E.~Dorokhov and L.~Tomio,
Phys.\ Rev.\ D {\bf 62} (2000) 014016; 
  M.~Praszalowicz and A.~Rostworowski,
  Acta Phys.\ Polon.\ B {\bf 34} (2003) 2699;
  F.~Bissey, \etal~ 
  Phys.\ Lett.\ B {\bf 547} (2002) 210;
  Phys.\ Lett.\ B {\bf 587} (2004) 189









\end{thebibliography}
\end{document}